\newlength{\pagewidth}  
\newlength{\pageheight} 
\newlength{\centremargin} 
\newlength{\outsidemargin} 
\newlength{\bottommargin} 
\newlength{\abstractindent}  
\newcommand{\mathtext}[1]{\mbox{$ #1 $}}
\renewcommand{\b}{\mathtext{b}}
\newcommand{\B}{\mathtext {B}}
\newcommand{\Kstar}{\mathtext {K^{*}}}
\newcommand{\K}{\mathtext K}
\newcommand{\psip}{\mathtext {\psi^{\prime}}}
\newcommand{\mup}{\mathtext {\mu^{+}}}
\newcommand{\mum}{\mathtext {\mu^{-}}}
\newcommand{\lp}{\mathtext {\ell^{+}}}
\newcommand{\lm}{\mathtext {\ell^{-}}}
\newcommand{\bsll}{\mathtext{b \rightarrow s \, \lp \lm}}
\newcommand{\BKll}{\mathtext{B \rightarrow K \, \lp \lm}}

\newcommand{\BKpsi}{\mathtext{B \rightarrow K \, \psi}}
\newcommand{\BKpsip}{\mathtext{B \rightarrow K \, \psi^{\prime}}}
\newcommand{\BKspsi}{\mathtext{B \rightarrow K^{*} \, \psi}}
\newcommand{\BKspsip}{\mathtext{B \rightarrow K^{*} \, \psi^{\prime}}}
\newcommand{\psill}{\mathtext{\psi \rightarrow \lp \lm}}

\newcommand{\bsmumu}{\mathtext{b \rightarrow s \, \mup \mum}}
\newcommand{\BKsmumu}{\mathtext{B \rightarrow \Kstar \, \mup \mum}}
\newcommand{\BKmumu}{\mathtext{B \rightarrow K \, \mup \mum}}
\newcommand{\bsgamma}{\mathtext{b \rightarrow s \gamma}}
\newcommand{\BKsgamma}{\mathtext{B \rightarrow \Kstar \gamma}}
\newcommand{\kaps}{\mathtext{\kappa^{*}}}
\newcommand{\mathd}{\mathtext{\rm d}}
\newcommand{\sigmunu}{\mathtext{\sigma^{\mu\nu}}}
\newcommand{\LamQCD}{\mathtext{\Lambda_{\rm QCD}}}
\newcommand{\IW}{\mathtext{\xi \left( v \cdot v^\prime \right)}}
\newcommand{\jou}[6]{#1 : #3  {\bf #4} (19#6) #5.}
\newcommand{\preprint}[3]{#1 : #3.}

\newcommand{\PRD}[5]{\jou{#1}{#2}{Phys.~Rev.~}{D#3}{#4}{#5}}

\newcommand{\PLB}[5]{\jou{#1}{#2}{Phys.~Lett.~}{B#3}{#4}{#5}}
\newcommand{\NPB}[5]{\jou{#1}{#2}{Nucl.~Phys.~}{B#3}{#4}{#5}}
\newcommand{\MPLA}[5]{\jou{#1}{#2}{Mod.~Phys.~Lett.~}{A#3}{#4}{#5}}

\newcommand{\refcite}[1]{ref.~\cite{#1}}
\newcommand{\refscite}[1]{refs.~\cite{#1}}
\newcommand{\figref}[1]{figure~\ref{fig:#1}}
\newcommand{\eqnref}[1]{equation~(\ref{eqn:#1})}
\documentstyle[12pt, epsf]{article}
\newlength{\abstractwidth}
\begin{document}
%
%
\begin{titlepage}
\noindent\hspace\fill UCLA/93/TEP/26
\newline
\vspace{0pt}\noindent\hspace\fill July 1993
\newline
\vspace{0pt}\noindent\hspace\fill (Bulletin Board : hep-ph/9307369)

\vfill
\begin{center}
\LARGE\bf Observing \mbox{$b \rightarrow s \, \mu^+ \mu^-$}
Decays at Hadron Colliders \\*[\fill]
\Large
Grant Baillie \\[3pt]
Department of Physics \\*
University of California, Los Angeles \\*
405 Hilgard Avenue \\*
Los Angeles, CA 90024--1547 \\*[3pt]
{\large (E-mail : {\tt baillie@physics.ucla.edu})} \\*[\fill]
%
{\large\bf Abstract} \\*[10pt]
\normalsize
\parbox{\abstractwidth}
{We examine the prospects for observing weak flavour-changing neutral
current (FCNC) decays of \B\ mesons at hadron colliders,
including effects of anomalous $WWZ$~vertices.
Since it is very difficult to measure the inclusive rate
$B \rightarrow X_s \, \lp \lm$ one should consider exclusive modes
such as \BKsmumu\ and \BKmumu. Even though this requires one to compute
hadronic matrix elements, we show that experimentally observable
quantities (ratios of decay rates) are not strongly parametrisation
dependent.
Some possibilities for reducing the theoretical uncertainties
from other experimental data are discussed.
} \\
\end{center}
\vfill
\end{titlepage}
\section{Introduction}

Because weak flavour-changing neutral currents (FCNC's)
are forbidden at tree level in the Standard Model Lagrangian, transitions
involving these currents are potentially sensitive tests of
electroweak radiative corrections. Beside their dependence on the as yet
undetermined top quark
mass~$m_t$, FCNC processes could provide a clear signal for ``new physics'',
stemming from the presence of non-Standard Model particles or couplings
in loop graphs.
For example, contributions to the decays \bsgamma\ and
\bsll\ have
been calculated in the context of supersymmetry~\cite{Bertolini},
two-Higgs models~\cite{Bertolini,Ciuchini}, anomalous
$WW\gamma$ couplings~\cite{WWgamma} and heavy fermions~\cite{InamiLim}.

Recently the CLEO collaboration, observing \mathtext{e^+e^-}~annihilation
at the
\mathtext{\Upsilon(4S)}~resonance, has detected the decay \BKsgamma\ 
\cite{CLEOpaper}; in addition an
upper bound on the branching fraction \mathtext{B (\bsgamma)}\ has been
claimed \cite{CLEOtalk}.
These results already provide interesting bounds on non-Standard Model
physics, which will be constrained further once the top quark is
discovered \cite{Rizzo}.
For \bsmumu, the UA1 collaboration has determined the experimental
upper bound to be $B(\bsmumu) < 5.0 \times 10^{-5}$, leaving room for
substantial
deviation from the standard model prediction $(6 - 8) \times 10^{-6}$
\cite{Ali}. Because of the large production cross-section for heavy quarks
at hadron colliders, an observation of these decays may be possible by
the D0 and CDF experiments, which have recently completed data
collection runs of proton-antiproton collisions at \mathtext{\sqrt{s}
= 1800}~GeV.  It is therefore important to make theoretical
predictions for these rare decays that take into account some of the
experimental difficulties associated with measuring them at hadron
machines.

The rate for \bsll\ can be computed quite straightforwardly once a
low-energy effective Lagrangian has been obtained by ``integrating
out'' heavy degrees of freedom.  Unfortunately, at hadron colliders it
is difficult to measure the corresponding inclusive branching ratio
\mathtext{B (B \rightarrow X_s \lp \lm)}, where $X_s$~denotes a final
state containing a strange (but not charm) quark. Because of the large
number of final state hadrons produced in any single event,
identifying a subset of particles as originating from a \B\ involves
reconstructing their total 4-momentum and invariant mass. The presence
of multiparticle states in the inclusive sum then makes it hard to
disentangle \B\ decay events from the general hadronic background.
Consequently, one is led to consider particular (or ``exclusive'')
decay modes, such as \mathtext{B \rightarrow \Kstar(892) \lp \lm}\ or
\BKll.  This is unpleasant from the theorist's point of view, because
one now needs to know the matrix elements of quark operators between
hadronic final states. At present, these cannot be computed from first
principles (i.e. from QCD), and various models and approximations have
to be used to estimate them. We will make a case for using some of the
results of heavy quark effective theory for the form factors here,
even though the $s$~quark cannot really be considered heavy. We will
estimate model-dependence by considering several different expressions
for the Isgur-Wise function, which parametrises the form factors.

For simplicity we choose to investigate only decay channels involving a pair
of muons. This is motivated partially by the fact that present detectors
at hadron colliders are more efficient at detecting muon pairs than electron
pairs in this energy regime. In addition, we note that the low-dimuon mass
region is very difficult to study experimentally, in part because the muons
tend to be less separated there, and also because of backgrounds (typically,
some kind of cut has to be imposed on muon transverse momentum --- see
\refcite{UA1}, for example).

We therefore consider the decays \BKsmumu\ and \BKmumu with dimuon
invariant mass above the \mathtext\psi\ peak in what follows. In
Section~2, we briefly review the amplitude for \bsll\ processes in the
Standard Model, taking into account QCD corrections and the
contribution of \mathtext{c \bar{c}}~pairs. After some discussion on
estimating hadronic matrix elements, we make some predictions within
the Standard Model.  As an example of how these predictions are
affected by ``new physics'', in particular new physics that cannot be
constrained by
\bsgamma\ decays, we consider the effect of non-Standard Model~$WWZ$
couplings in Section~4.  The results will be discussed in
Section~\ref{sec:Discussion}, and finally, we draw conclusions in
Section~\ref{sec:Conclusions}.

\section{The process \bsll }
The amplitude for the decay~\bsll\ can be written in the form
\begin{equation}
  {\cal M} = {4 G_F \over \sqrt{2}}
             {\alpha(m_b) \over 4 \pi x_{\scriptstyle W}}
             V^{*}_{ts} V_{tb}
             \left\{
                A \: \bar{s} \gamma^\mu L b \: \bar\ell \gamma_\mu L \ell
              + B \: \bar{s} \gamma^\mu L b \: \bar\ell \gamma_\mu R \ell
              + x_W F \: {i \over q^2} m_b \,
                      \bar{s} \sigmunu q_\nu R b
                              \: \bar\ell \gamma_\mu \ell
            \right\},
\label{eqn:bsll}
\end{equation}
where~$\alpha(m_b) \approx 1 / 132.7$ is the QED running coupling evaluated
at the \b~scale,
$G_F$~is the Fermi weak decay constant,
$x_W = \sin^2 \theta_W$ ($\theta_W$~is the
Weinberg angle), $V$~is the Kobayashi-Maskawa (KM) matrix, $s$~and~$b$ are
quark spinors and
$R$~and~$L$ are the projection operators $(1 + \gamma^5) / 2$ and
$(1 - \gamma^5) / 2$, respectively. We have defined~$q^\mu$ to be the
total outgoing 4-momentum of the final $\lp \lm$~pair
in the decay \bsll. In addition, in the $F$~term, a contribution proportional
to $m_s$ has been dropped.
Throughout, we use the conventions
$\sigma_{\mu\nu} = i/2 \left[ \gamma_\mu, \gamma_\nu \right]$ and
$\epsilon_{0123} = + 1$.

\begin{figure}
\begin{center}
\leavevmode
\epsffile{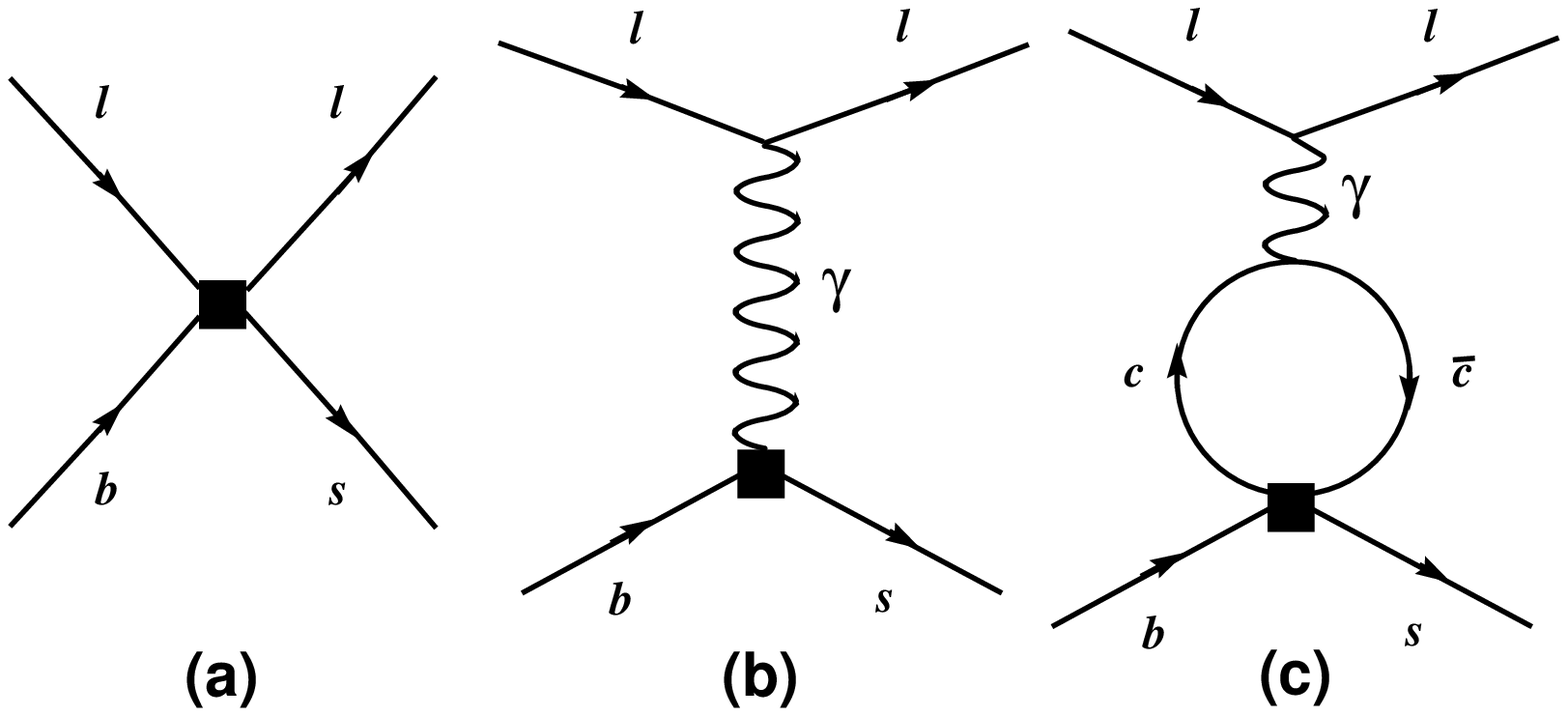}
\end{center}
\caption{
Contributions to the amplitude for \bsll. The ``black boxes''
denote higher-dimension operators induced by integrating out the top,
$W$~and~$Z$ at a scale~$\mu = M_W$.}
\label{fig:bsll}
\end{figure}

In an effective field theory approach, $A$, $B$~and~$F$ arise
from the graphs of \figref{bsll}.
(Note that the contribution
of $u$-quark graphs is KM suppressed; we make the usual approximation
$V^{*}_{ts} V_{tb} \approx - V^{*}_{cs} V_{cb}$,
$V^{*}_{us} V_{ub} \approx 0$.)
The $\bar{s}b \, \bar\ell \ell$~terms of \figref{bsll}(a) appear
in the low-energy effective Lagrangian as a result of
integrating out $Z$~exchange and $W$~box diagrams at $\mu\sim M_W$.
Also, $\bar{s} b \gamma$ vertices contribute via
\figref{bsll}(b).
In particular, the $1/q^2$~pole
in front of the $F$~term above is a result of there being an on-shell
photon intermediate state at~$q^2 = 0$ (the coefficient~$F$ also appears
in the rate for \bsgamma).
Finally, the sum of the two graphs represented by \figref{bsll}(c) can
be evaluated as in \refcite{bsll-longd}, where the charm loop is expressed
as a dispersion integral which receives
continuum (essentially free quark) and $c \bar{c}$ resonance contributions.
The latter,
because of the narrow widths involved, amount to
adding Breit-Wigner terms for the
\mathtext{\psi\,(3097)}\ and \mathtext{\psip \,(3685)}\ to both
$A$~and~$B$:
\begin{equation}
A \: (B) \quad \rightarrow \quad A \: (B) + (C_9 + 3 C_{10}) \,
{3 \pi x_W \over \alpha^2(m_b)} \,
{
M_\psi \Gamma ( \psill )
\over
q^2 - M_\psi^2 + i M_\psi \Gamma_\psi
}
\: + \: (\psi \rightarrow \psi^\prime)
\label{eqn:BreitWigner}
\end{equation}
where
\begin{equation}
C_9 + 3 C_{10} = 
  2 \left\{ \alpha_s \, (m_b) / \alpha_s \, (M_W) \right\}^{-6/23}
  - \left\{ \alpha_s \, (m_b) / \alpha_s \, (M_W) \right\}^{12/23}
\label{eqn:bsccQCD}
\end{equation}
is a QCD-corrected Wilson coefficient evaluated at the \b~scale.
In general, then, the coefficients~$A$, $B$~and~$F$ depend on
$x_t \equiv m_t^2 / M_W^2$
(a result of integrating out heavy degrees of freedom
at the $W$~scale), and also contain QCD correction terms (computed by
running Wilson coefficients in the effective Lagrangian down from
$\mu = M_W$ to $\mu = m_b$) and both short- and long-distance contributions
from charm loops. Explicit expressions can be found in
\refcite{Toronto:summary}; detailed calculations are to be found in the
references therein. In terms of their coefficient
functions~$A_{1,2,3}$ of~\cite{Toronto:summary}\ we have

\begin{eqnarray}
A & = & x_W (A_1 + A_3) \\
B & = & x_W A_1 \\
F & = & -2 A_2
\end{eqnarray}

We also note that, for values of the top quark mass in the range
100--200 GeV, the coefficient~$A$ (excluding the charm contributions)
is a good deal larger than $B$~or~$x_W F$. For example, at~$m_t =
150$~GeV we find $A = 1.58$, $B = -0.0737$ and $x_W F = 0.144$ when
$\Lambda_{\rm QCD}^{(4)} = 150$~MeV.

\section{Exclusive rare \B\ decays}
To compute the rates for \BKsmumu~and~\BKmumu\ in the spectator quark
approximation, one needs to know the matrix elements of~$\bar{s}
\gamma^\mu L b$ and $\bar{s} i \sigmunu q_\nu R b$ sandwiched
between the initial and final hadronic states. In this paper, we
assume that these matrix elements have the form prescribed by heavy
quark effective theory (HQET) \cite{IsgurWise}:
\begin{equation}
\begin{array}{rcl}
\displaystyle
  \left\langle
    K(v^\prime)
  \right|
  \bar{s} \gamma^\mu L b
  \left|
    B(v)
  \right\rangle
& = &
\displaystyle
  - {1 \over 2} \sqrt{M_B M_K} \; \IW \;
    \left( v + v^\prime \right)^\mu \\[10pt]
\displaystyle
  \left\langle
    K(v^\prime)
  \right|
  \bar{s} i \sigmunu q_\nu R b
  \left|
    B(v)
  \right\rangle
& = &
\displaystyle
  - {1 \over 2} \sqrt{M_B M_K} \; \IW \;
    \left(
       v^{\prime\mu} v^\nu
     - v^\mu v^{\prime\nu}
    \right)  q_\nu
\end{array}
\label{eqn:Kformfactors}
\end{equation}
and
\begin{equation}
\begin{array}{rcl}
\displaystyle
  \left\langle
    \Kstar(v^\prime, \epsilon)
  \right|
    \bar{s} \gamma^\mu L b
  \left|
    B(v)
  \right\rangle
& = &
\displaystyle
  {1 \over 2} \sqrt{M_B M_{K^{*}}} \; \IW \; \times
\\[10pt]
& &
\quad
\displaystyle
    \left(  i \epsilon^{\mu\alpha\beta\gamma} \;
              \epsilon^{*}_\alpha \;
              v_\beta \;
              v^\prime_\gamma
          + (1 + v \cdot v^\prime) \; \epsilon^{*\mu}
          - v \cdot \epsilon^{*} \; v^{\prime\mu}
    \right) \\[10pt]
\displaystyle
  \left\langle
    K^{*}(v^\prime)
  \right|
  \bar{s} i \sigmunu q_\nu R b
  \left|
    B(v)
  \right\rangle
& = &
\displaystyle
  {1 \over 2} \sqrt{M_B M_{K^{*}}} \; \IW \; \times
\\[10pt]
& & 
\quad
\displaystyle
  \left(
     i \epsilon^{\mu\nu\alpha\beta} \;
       \epsilon^{*}_\alpha \;
       (v + v^\prime)_\beta
     + (v + v^\prime)^\mu \epsilon^{*\nu}
     - (v + v^\prime)^\nu \epsilon^{*\mu}
  \right)  q_\nu .
\end{array}
\label{eqn:Kstarformfactors}
\end{equation}

Here $\epsilon$~is the \Kstar\ polarisation vector, $v$~and~$v^\prime$
are the \B\ and \mathtext{K^{(*)}} 4-velocities (so that 4-momenta are
given
by~$p_B^\mu = M_B v^\mu$,
$p_{K^{(*)}}^\mu = M_{K^{(*)}} v^{\prime\mu}$)
and \IW\ is the Isgur-Wise function. Strictly speaking, the
4-vector~$q^\mu$ is the difference between the $b$~quark and $s$~quark
momenta, but we make the usual identification
$q^\mu = p_B^\mu - p_{K^{(*)}}^\mu$.

Of course, one should be suspicious of using the heavy quark method in
this case, because corrections to its predictions are expected to be
of order~$\LamQCD / m_Q$, which is not a particularly small number for
the strange quark.  In the \Kstar~case, a general Lorentz-invariant
decomposition of the left-hand sides of~\eqnref{Kstarformfactors}\
would involve seven different functions of~$q^2$, which all turn out
to be related to the Isgur-Wise function~\IW\ by the heavy quark
symmetry when \mathtext{ \: m_s \; , \; m_b \gg \LamQCD}.  In spite of
the fact that \mathtext{ \: m_s \not \gg \LamQCD }, it is shown in
\refscite{ODT:heavylight,ODT:BKStarGamma} that these relations amongst
form factors coninue to hold to about~10\%, mainly as a result of the
heaviness of the \b~quark, and also the fact that the \Kstar~meson is
in some sense ``weakly bound''.

Note that one can test some of the heavy-quark
relations amongst \mathtext{\bar{s} \gamma^\mu L b}~form factors
by measuring the polarisation of the \Kstar~meson in the decays
\BKspsi and \BKspsip. Even though this leaves open the possibility of
deviations in the remaining \mathtext{\bar{s} \sigmunu b} matrix
elements, the two sets of form factors can be related by the plausible
assumption that the \b~quark is static within the \B~meson
\cite{ODT:heavylight}. 

The \K~meson, on the other hand, is a relativistic bound state, and it
might not be appropriate to apply the constituent quark model analysis
of~\cite{ODT:heavylight,ODT:BKStarGamma} here. Nevertheless, if one
performs a general decomposition of the matrix elements
in~\eqnref{Kformfactors}, one finds that only two of the three
resulting form factors contribute to \BKmumu. In addition, because
\mathtext{A \gg B {\rm\ or\ }x_W F},
one of the two will dominate this decay. (Note that there are no
\mathtext{q^2 = 0}~poles in the decay rate to enhance the contribution
of~$F$: these would be a result of an on-shell intermediate
\mathtext{B \rightarrow K \gamma} transition, which is forbidden by
angular momentum conservation). Consequently, deviations from the
heavy quark relations amongst \K~form factors are unlikely to have a
great effect on the rate.

It should be pointed out that extending the heavy quark spin-flavour
symmetry to the $s$~quark case could be dubious.  In other words, one
cannot assume that the function~$\xi$ above is related to the
function~$\xi$ determined from \b~and~$c$ decays (flavour symmetry);
nor can one really say that the functions~$\xi$ in equations
(\ref{eqn:Kstarformfactors})~and~(\ref{eqn:Kformfactors}) are the same
(spin symmetry).  (However, we shall continue to refer to the two
functions $\xi_K$~and~$\xi_{K^*}$ generically as~$\xi$, unless we need
to distinguish between the two.) Essentially, we are using the heavy
quark form of the matrix elements as a convenient ansatz which appears
to hold to a good degree of accuracy.

Another property of HQET is that at the zero-recoil point, where $v
\cdot v^\prime = 1$, the Isgur-Wise function satisfies $\xi \; (1) =
1$. We are reluctant to assume this is the case in
\mathtext{b \rightarrow s}~transitions --- for example, one can note that
$m_s$, $M_K$ and $M_{K^{*}}$ are all supposed to be degenerate in the
heavy quark limit. In constituent quark models one finds that the
above normalisation condition holds in the case of large quark masses.
However, an explicit computation of form factors in the na\"\i ve
quark model of \refcite{ODT:heavylight} gives $\xi \; (1) \approx 0.7$
in the \mathtext{B \rightarrow K}~case. (This is not surprising, as
$1/m_Q$~corrections are known to be large for
pseudoscalar~$\rightarrow$~pseudoscalar transitions). Consequently, we
avoid issues of normalisation of our form factors by taking ratios of
decay rates. This is also desirable from the experimental point of
view, as some uncertainties, like luminosity and detector efficiency,
then tend to cancel.

In view of our earlier comments about the difficulty of observing the
small~$q^2$ region, it makes sense to divide out the
\mathtext\psi~peak, and make predictions for the quantities

\begin{equation}
R^{(*)} = 
{
\displaystyle
\Gamma_1 \left(B \rightarrow K^{(*)} \mup \mum \right)
\over
\displaystyle
\Gamma_2 \left(B \rightarrow K^{(*)} \mup \mum \right)
} 
\end{equation}

where $\Gamma_1$~is the contribution of the region of phase space with
\mathtext{\hat{s}\equiv q^2 / M_B^2}~above the $\psi$~peak,
with the \psip\ excluded,
while $\Gamma_2$ is the contribution of the $\psi$~peak itself. Specifically,
we will assume

\begin{eqnarray}
\Gamma_1 &=&
  \left( \int_{0.35}^{0.48} \mathd \hat{s}
       + \int_{0.50}^{ {\hat{s}}_{\rm max}} \mathd \hat{s}
  \right)
  \;
  {\mathd \over \mathd \hat{s}} \,
  \Gamma
\label{eqn:Gamma1}
\\[15pt]
\Gamma_2 &=&
  \int_{0.34}^{0.35} \mathd \hat{s}
  {\mathd \over \mathd \hat{s}} \,
  \Gamma \: ,
\label{eqn:Gamma2}
\end{eqnarray}

since \mathtext{ {\hat{s}}_\psi \approx 0.344 }
and \mathtext{ {\hat{s}}_{\psi^\prime} \approx 0.487 }.
Here the differential decay widths can be computed
from equations~(\ref{eqn:bsll}),
(\ref{eqn:Kformfactors})~and~(\ref{eqn:Kstarformfactors}):
\begin{equation}
\begin{array}{rcl}
\displaystyle
  {\mathd \over \mathd \hat{s}} \,
  \Gamma \, (\BKmumu)
& = &
\displaystyle
  {G_F^2 M_B^5 \over 192 \pi^3} \,
  \left| V^*_{ts} V_{tb} \right|^2 \,
  \left( {\alpha(m_b) \over 4 \pi x_W} \right)^2 \,
  2 \kappa^2 \, \xi^2 (y) \, \left( y^2 - 1 \right)^{3/2}
 \quad \times
\\[10pt]
& &
\displaystyle \phantom{+}
  \left\{ \vphantom{| \over |}
    - 2 \, (\kappa + 1) \, F \, \mbox{\rm Re} \, (A + B) \,
    + (\kappa + 1)^2 \, \left( \left| A \right|^2 + \left| B \right|^2 \right)
    + 2 \, F^2 \vphantom{| \over \|}
  \right\}
\end{array}
\label{eqn:KRate}
\end{equation}
and
\begin{equation}
\begin{array}{rcl}
\displaystyle
  {\mathd \over \mathd \hat{s}} \,
  \Gamma \, (\BKsmumu)
& = &
\displaystyle
  {G_F^2 M_B^5 \over 192 \pi^3} \,
  \left| V^*_{ts} V_{tb} \right|^2 \,
  \left( {\alpha(m_b) \over 4 \pi x_W} \right)^2 \,
  2 \kaps^2 \, (1 + y) \, \xi^2 (y) \, \left(y^2 - 1\right)^{1/2} \quad \times
\\[10pt]
& &
\displaystyle \phantom{+}
  \left\{ \vphantom{| \over |}
    2 \, F \, \mbox{\rm Re} \, (A + B) \,
    \left( (5 \kaps - 1) + (\kaps - 5)^{\phantom{1}} y \right)
  \right.
\\[10pt]
& &
\displaystyle \phantom{+}
  \left. \, + \,
    {2 F^2 \over \hat{s}} \,
    \left(
       (\kaps^2 - 8 \kaps + 1) +
       (5 \kaps^2 - 2 \kaps + 5) \; y - 2 \kaps y^2
    \right)
  \right.
\\[10pt]
& &
\displaystyle \phantom{+}
  \left. \, + \,
    \left( \left| A \right|^2 + \left| B \right|^2 \right) \,
    \left( (\kaps - 1)^2 + (5 \kaps^2 - 2\kaps + 5) \; y - 8\kaps y^2 \right)
    \vphantom{| \over |}
  \right\}.
\end{array}
\label{eqn:KStarRate}
\end{equation}

We have defined
\mathtext{\kappa \equiv M_K / M_B},
\mathtext{\kaps \equiv M_{K^{*}} / M_B},
and
\mathtext{y \equiv v \cdot v^\prime 
            = 1/2 \left( (1 - \hat{s}) / \kappa^{(*)}
            + \kappa^{(*)} \right)}.

It should be noted that~$\Gamma_2$ is essentially proportional to the square
of the QCD coefficient in \eqnref{bsccQCD}. However, as pointed out in
\refcite{Kuhn-Ruckl}, because of an accidental cancellation, this quantity is
highly sensitive to the value of~\LamQCD. For example, we find that, for the
\Kstar~transition, \mathtext{\Gamma_2} changes by a factor of about two as
\mathtext{\Lambda^{(4)}_{QCD}}~varies from 150 to 300 MeV.
Following \cite{Kuhn-Ruckl},
we will replace this coefficient in the Breit-Wigner amplitudes
by its QCD-uncorrected value of 1, which gives a reasonably good agreement
with the measured~\mathtext{B \rightarrow X \psi}~inclusive branching ratio.

Finally, we choose various parametrisations of the function~$\xi$ above.
First of all, we consider the simple monopole and exponential expressions
\begin{equation}
\xi_I \, (y) = {w_0^2 / 2 \over w_0^2 / 2 - 1 + y}
\label{eqn:IWfirst}
\end{equation}
\begin{equation}
\xi_{II} \, (y) = \exp \alpha \, (1 - y),
\end{equation}

taking for $\alpha$~and~$w_0$ the values determined
from $D \rightarrow K \ell \nu$~decays in~\cite{Ali-Mannel}:
\mathtext{\alpha \approx 0.5} and \mathtext{w_0 \approx 1.8}.

Also, we consider the two forms given in \refcite{Neubert-review}:
\begin{equation}
\xi_{III} \, (y) = \left( {2 \over y + 1} \right)^{2 \rho^2}
\end{equation}
\begin{equation}
\xi_{IV} \, (y) = {2 \over y + 1}
                    \exp \left(-\beta {y - 1 \over y + 1} \right),
\label{eqn:IWlast}
\end{equation}

with $\beta$~and~$\rho$ computed from \mathtext{B \rightarrow D^{*} \ell \nu}
to be \mathtext{\beta \approx 1.84} and \mathtext{\rho \approx 1.14}.

\begin{figure}
\begin{center}
\leavevmode
\epsffile{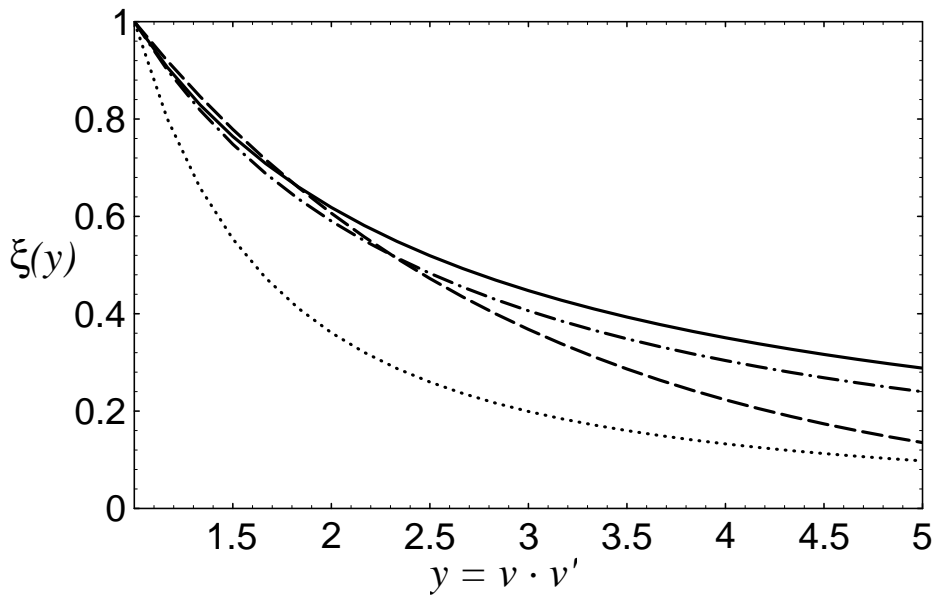}
\end{center}
\caption{Plot of the four different parametrisations of the
Isgur-Wise function~\mathtext{ \xi_i (y) }.
\mathtext{i =}~I is given by solid lines,
II~by dashed, III~by dotted and IV~by dot-dashed lines.
See equations
(14)--(17).
}
\label{fig:IsgurWise}
\end{figure}

\begin{figure}
\begin{center}
\leavevmode
\epsffile{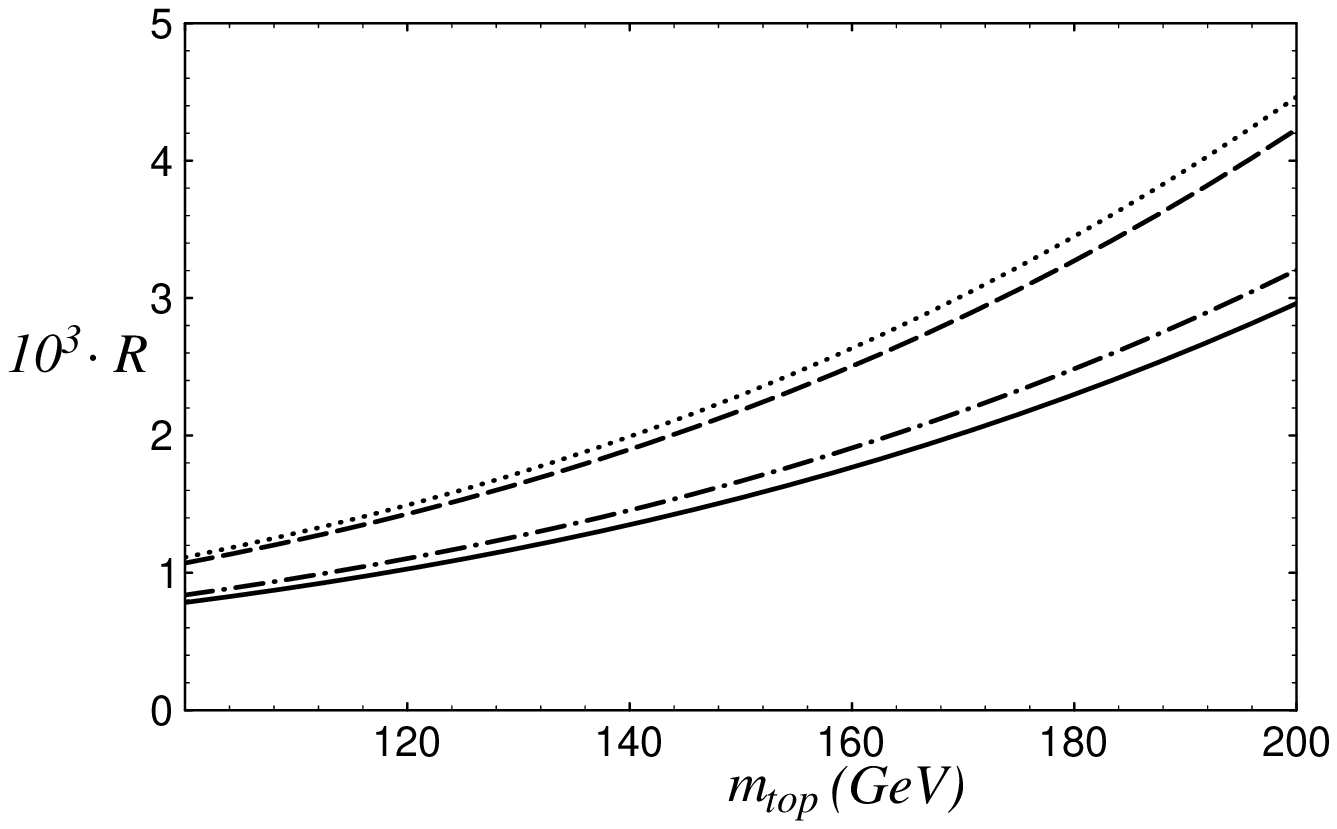}
\end{center}
\caption{Plot of the quantity~$R$ versus~$m_t$,
for various parametrisations of the
Isgur-Wise function.
}
\label{fig:BKmt}
\end{figure}

The above four
functions are plotted in \figref{IsgurWise}.
Although it might seem strange to use fits to both
\mathtext{D \rightarrow K \ell \nu} and
\mathtext{B \rightarrow D^{*} \ell \nu}
decays, this at least gives some kind of idea of uncertainties due to
the strange quark mass not being small.
Plots of $R$~and~$R^{*}$, for top quark masses between 100 and 200 GeV,
are presented in figures \ref{fig:BKmt}~and~\ref{fig:BKsmt}. We defer
discussion of these results to Section~\ref{sec:Discussion}.

\section{Effect of non-SM $WWZ$ couplings}

\begin{figure}
\leavevmode
\epsffile{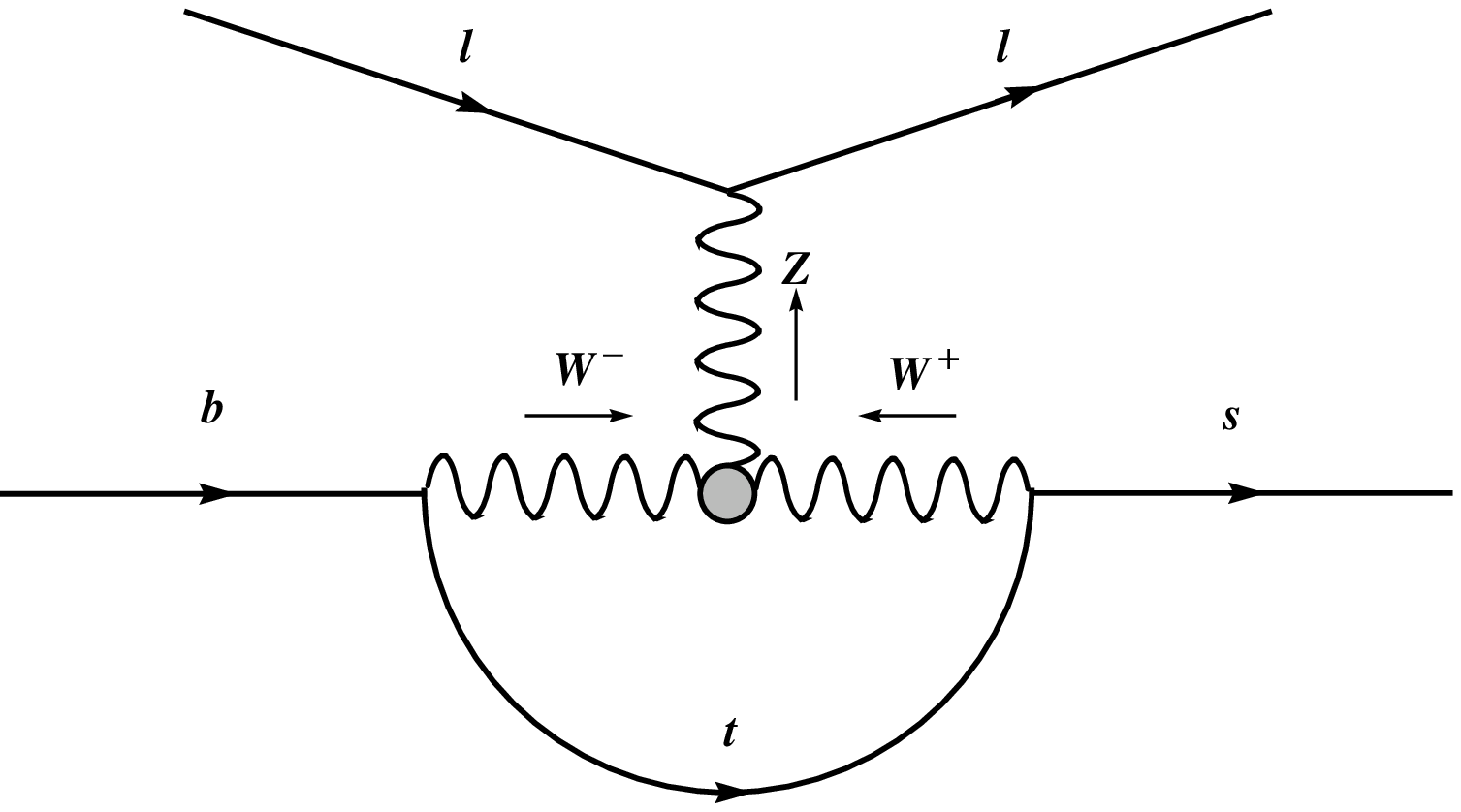}
\caption{Contribution of anomalous $WWZ$~couplings to \bsll }
\label{fig:bsllWWZ}
\end{figure}

\begin{figure}
\begin{center}
\leavevmode
\epsffile{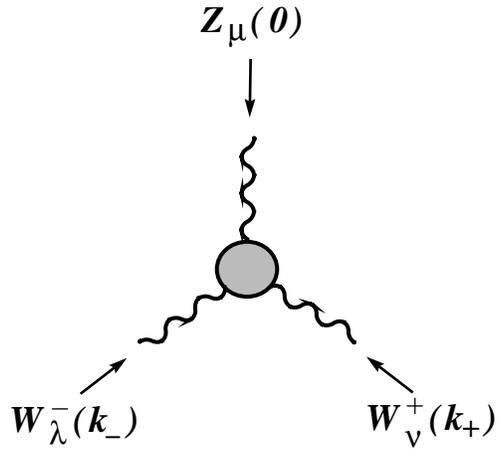}
\end{center}
\caption{Definitions for the Feynman rule for the
non-Standard Model part of the $WWZ$~vertex, at zero $Z$~momentum}
\label{fig:WWZ}
\end{figure}

The contribution of a non-Standard Model $WWZ$~vertex to the \bsll~amplitude
is shown in
\figref{bsllWWZ}. At a scale~$\mu = M_W$, we can assume all external
lines have zero 4-momentum. In this case, we can ignore derivatives of the
$Z$~field in the anomalous vertex, so that the full Lorentz- and 
$U(1)$-invariant vertex of \refcite{WWV} reduces to two terms:

\begin{displaymath}
{\cal L}_{\rm WWZ} =
    -i e \cot \theta_W \; Z_\mu
     \left\{
        \;
        g_1^Z
        \left(
           W_\nu W^{\dag \; \mu \nu}
         - W^{\dag}_\nu W^{\mu \nu}
        \right)
      - i g_5^Z \epsilon^{\mu \nu \rho \sigma}
                W^{\dag}_\nu
                \mathop{\partial_\rho}^\leftrightarrow
                W_\sigma
        \;
     \right\}
\end{displaymath}

where~$W^\mu$ is the $W^{-}$~field, and
$A \; {\displaystyle\mathop{\partial_\rho}^\leftrightarrow} \, B \equiv
A \left( \partial_\rho B \right) - \left( \partial_\rho A \right) B$. In the
Standard Model, $g_1^Z = 1$ and $g_5^Z = 0$, so the
Feynman rule for the (non-SM) vertex of \figref{WWZ} is:
\begin{displaymath}
-i e \cot \theta_W
    \left\{
      \Delta g_1^Z
      \left[ \left( k_+ - k_- \right)^\mu g^{\nu\lambda}
           + k_-^\nu g^{\mu \lambda}
           - k_+^\lambda g^{\mu \nu}
      \right]
      + i g^Z_5 \epsilon^{\nu \lambda \rho \mu} 
                \left( k_- - k_+ \right)_\rho
    \right\}.
\end{displaymath}

In an effective Lagrangian formalism, we need to compute the graph of
\figref{bsllWWZ}, which amounts to computing the non-Standard Model
\mathtext{\bar{s} b Z}~vertex at zero external momentum.
Working for the moment in $R_\xi$~gauge, we find the
\mathtext{b \rightarrow s Z} amplitude to be
\begin{equation}
\begin{array}{rl}
\displaystyle
{g^3 \cos \theta_W \over 2} \,
V^{*}_{ts} V_{tb} \,
\int
{\mathd^4 k \over \left( 2 \pi \right)^4} \,
& \displaystyle
{-\bar{s} \; \gamma_\alpha \not k \, \gamma_\beta L \; b \over 
 \left( k^2 - M_W \right)^2}
\;
\left( {1 \over k^2 - m_t^2} - {1 \over k^2} \right)^{\vphantom 2} 
\:
D^{\beta\lambda}_\xi \, (k) \; D^{\alpha\nu}_\xi \, (k)
\times
\\[10pt]
&
\left(
\Delta g_1^Z 
\left(
  k^\nu g^{\mu \lambda} - 2 k^\mu g^{\lambda \nu} + k^\lambda g^{\mu\nu}
\right)
+ 2 i g_5^Z \epsilon^{\nu\lambda\rho\mu} k_\rho
\right)
\end{array}
\label{eqn:bsZ}
\end{equation}
with
\mathtext{
D^{\alpha\beta} \, (k) =
g^{\alpha \beta} - (1 - \xi) k^\alpha k^\beta / (k^2 - \xi M_W^2)
}. Note that here we have used GIM cancellation to subtract off the
$m_t$-independent portion \cite{InamiLim}.
In the above, it is clear that the $g^Z_5$~contribution is
gauge-independent --- the longitudinal terms in the $W$~propagators don't
contribute --- and finite. However, the
$\Delta g^Z_1$~term does depend on~$\xi$. This term should therefore be
computed in the physical (unitary) gauge, obtained in the limit
\mathtext{\xi \rightarrow \infty}. However, in this limit the integral
is logarithmically divergent. (Note that computing this graph with a Standard
Model $WWZ$~vertex also gives a divergent result in unitary gauge, but
gauge invariance leads to a ``miraculous'' cancellation between divergences
in physical amplitudes.)
We can regulate this divergence via the prescription
\begin{equation}
\xi \rightarrow \Lambda^2 / M_W^2 \gg 1,
\end{equation}
with $\Lambda$~a cutoff. The integral
in (\ref{eqn:bsZ}) can now be evaluated explicitly, giving
\begin{equation}
{g^3 \cos \theta_W \over 2} \,
V^{*}_{ts} V_{tb} \,
{i \over 16 \pi^2} \,
\bar{s} \gamma_\mu L b \,
\left( \Delta g_1^Z F_1 (x_t, \Lambda)
     + g_5^Z F_5 (x_t )
\right)
\end{equation}
where
\begin{eqnarray}
x_t &\equiv& m_t^2 / M_W^2, \\[10pt]
F_1 (x_t, \Lambda)
  &=& 
    {3 \over 2} x_t
    \left(
       \log {\displaystyle \Lambda^2 \over M_W^2} +
       {1 \over 1 - x_t} +
       {2x_t - x_t^2 \over \left( 1 - x_t \right)^2} \log x_t 
    \right), \\[10pt]
F_5 (x_t)
  &=&
    -{3 x_t \over 1 - x_t}
    \left( 1 + {\displaystyle x_t \log x_t \over 1 - x_t} \right).
\end{eqnarray}

Thus the contribution of the anomalous vertex to the amplitude
of~\eqnref{bsll} is
\begin{equation}
\begin{array}{rl}
{\cal M} =
&
\displaystyle
  {4 G_F \over \sqrt{2}} \,
  {\alpha(m_b) \over 4 \pi x_{\scriptstyle W}} \,
  V^{*}_{ts} V_{tb} \,
  (1 - x_W) \,
  \left( \Delta g_1^Z F_1 (x_t, \Lambda)
       + g_5^Z F_5 (x_t )
       \right)
  \times
\\
&
  \quad
  \left(
    -(x_W - {1 \over 2})  \,
    \bar{s} \gamma^\mu L b \: \bar\ell \gamma_\mu L \ell
    -x_W \,     \bar{s} \gamma^\mu L b \: \bar\ell \gamma_\mu R \ell
  \right).
\end{array}
\label{eqn:bsll-anomalous}
\end{equation}

It turns out that four-Fermi operators of the above type do not scale
below~\mathtext{\mu = M_W}, because \mathtext{\bar{s} \gamma^\mu (\gamma^5) b}
are currents of a softly-broken symmetry. Consequently, we can ignore
QCD corrections to \eqnref{bsll-anomalous}.

\begin{figure}
\begin{center}
\leavevmode
\epsffile{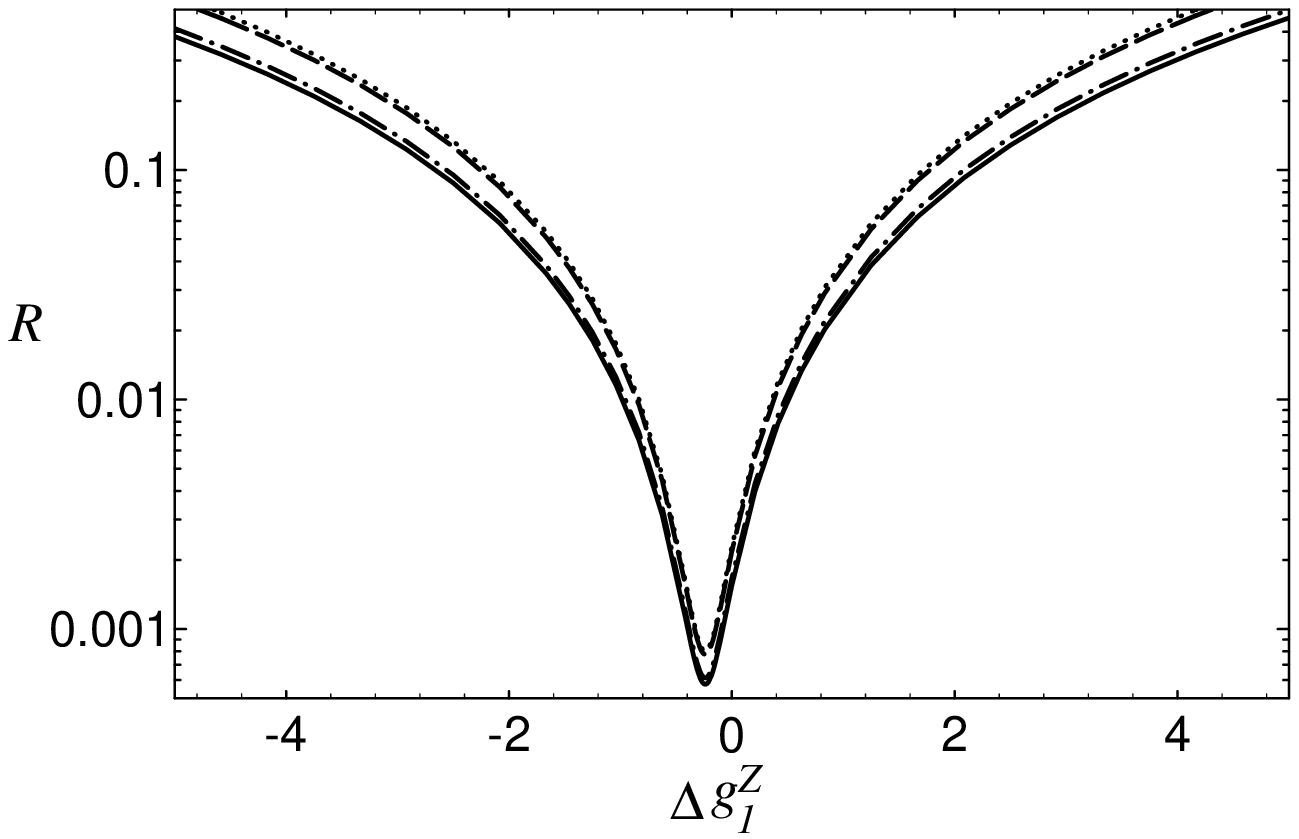}
\end{center}
\caption{Plot of the quantity~$R$ versus~$\Delta g_1^Z$,
for various parametrisations of the
Isgur-Wise function.
}
\label{fig:BKg1}
\end{figure}

\begin{figure}
\begin{center}
\leavevmode
\epsffile{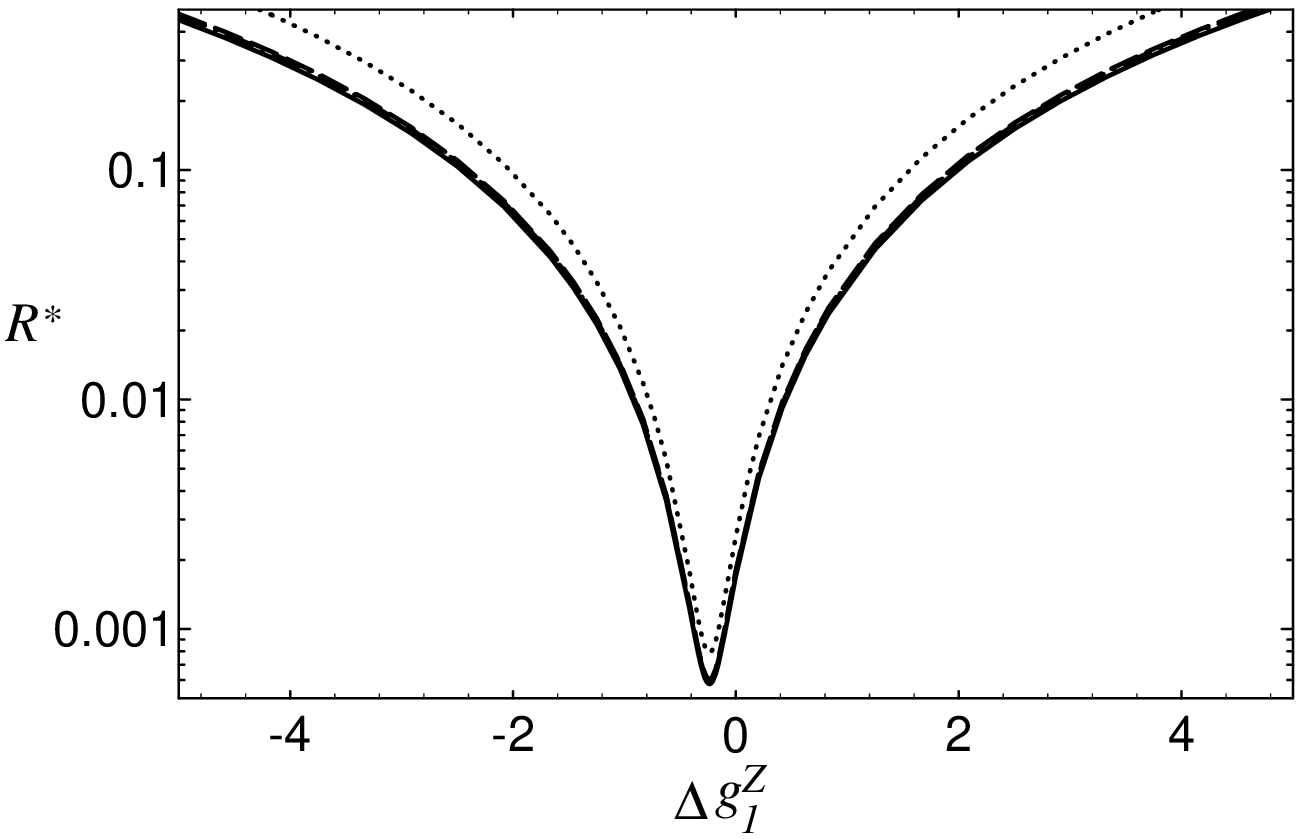}
\end{center}
\caption{Plot of the quantity~$R^{*}$ versus~$\Delta g_1^Z$,
for various parametrisations of the
Isgur-Wise function.
}
\label{fig:BKsg1}
\end{figure}

\begin{figure}
\begin{center}
\leavevmode
\epsffile{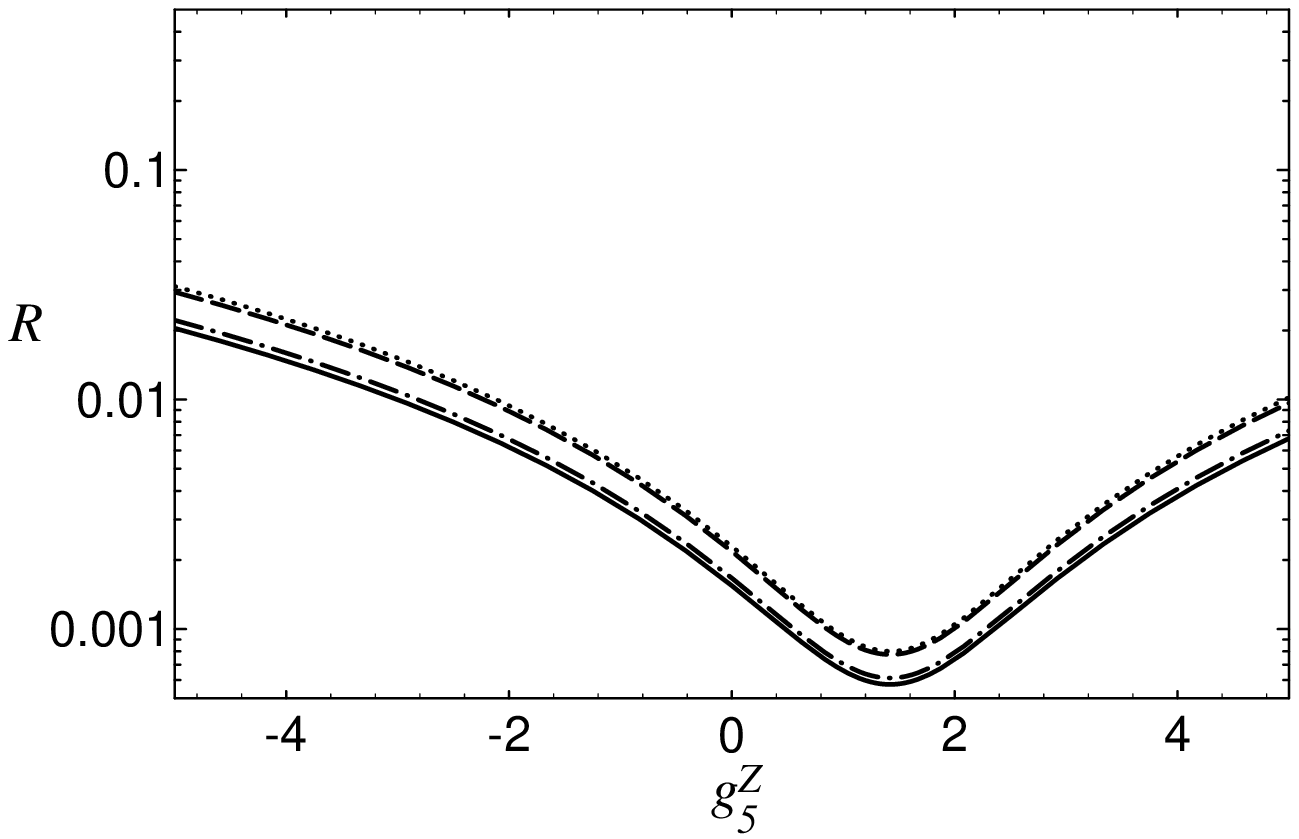}
\end{center}
\caption{Plot of the quantity~$R$ versus~$g_5^Z$,
for various parametrisations of the
Isgur-Wise function.
}
\label{fig:BKg5}
\end{figure}

\begin{figure}
\begin{center}
\leavevmode
\epsffile{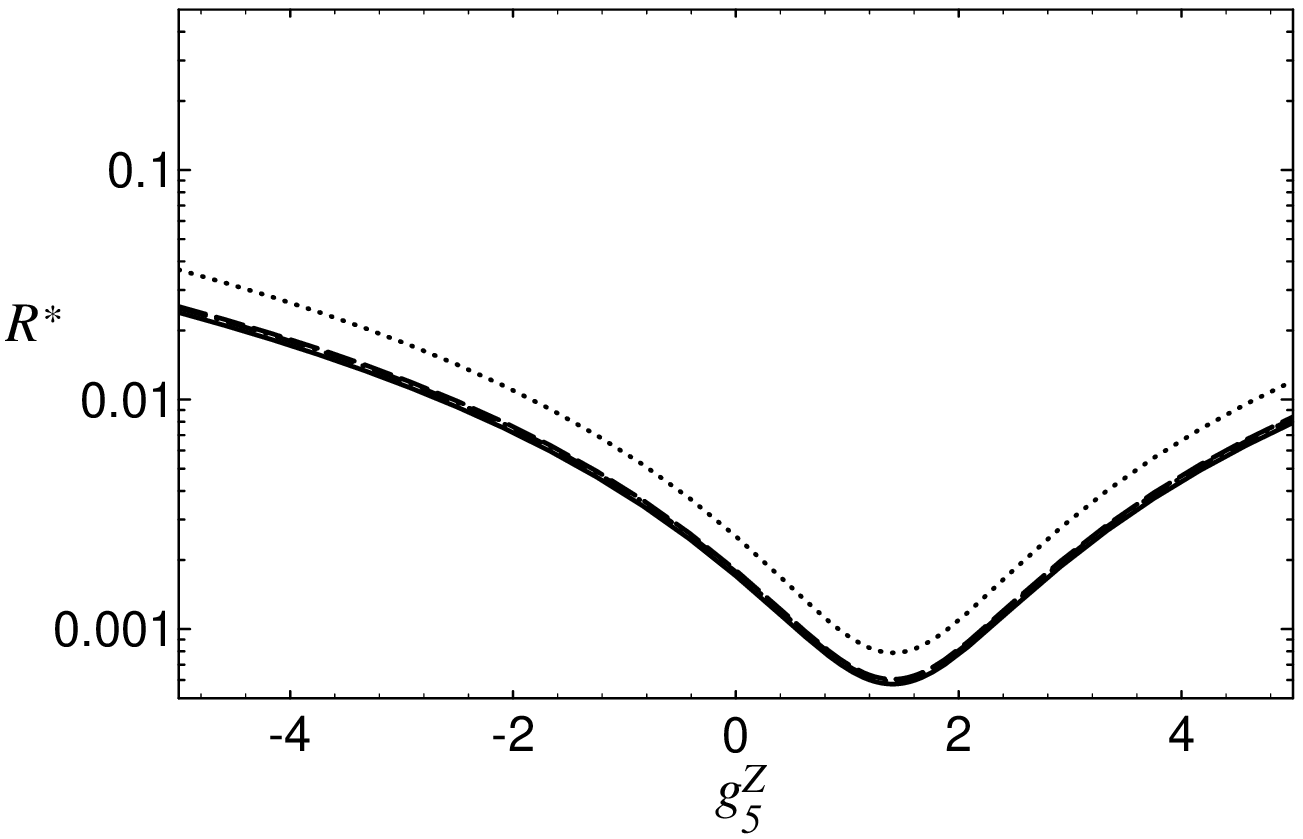}
\end{center}
\caption{Plot of the quantity~$R^{*}$ versus~$g_5^Z$,
for various parametrisations of the
Isgur-Wise function.
}
\label{fig:BKsg5}
\end{figure}

In figures~\ref{fig:BKg1}--\ref{fig:BKsg5} we fix \mathtext{m_t = 150}~GeV
and \mathtext{\Lambda = 1}~TeV, and plot $R$~and~$R^{*}$ as functions of
$\Delta g_1^Z$~and~$g_5^Z$.

\section{Discussion}
\label{sec:Discussion}
\begin{figure}
\begin{center}
\leavevmode
\epsffile{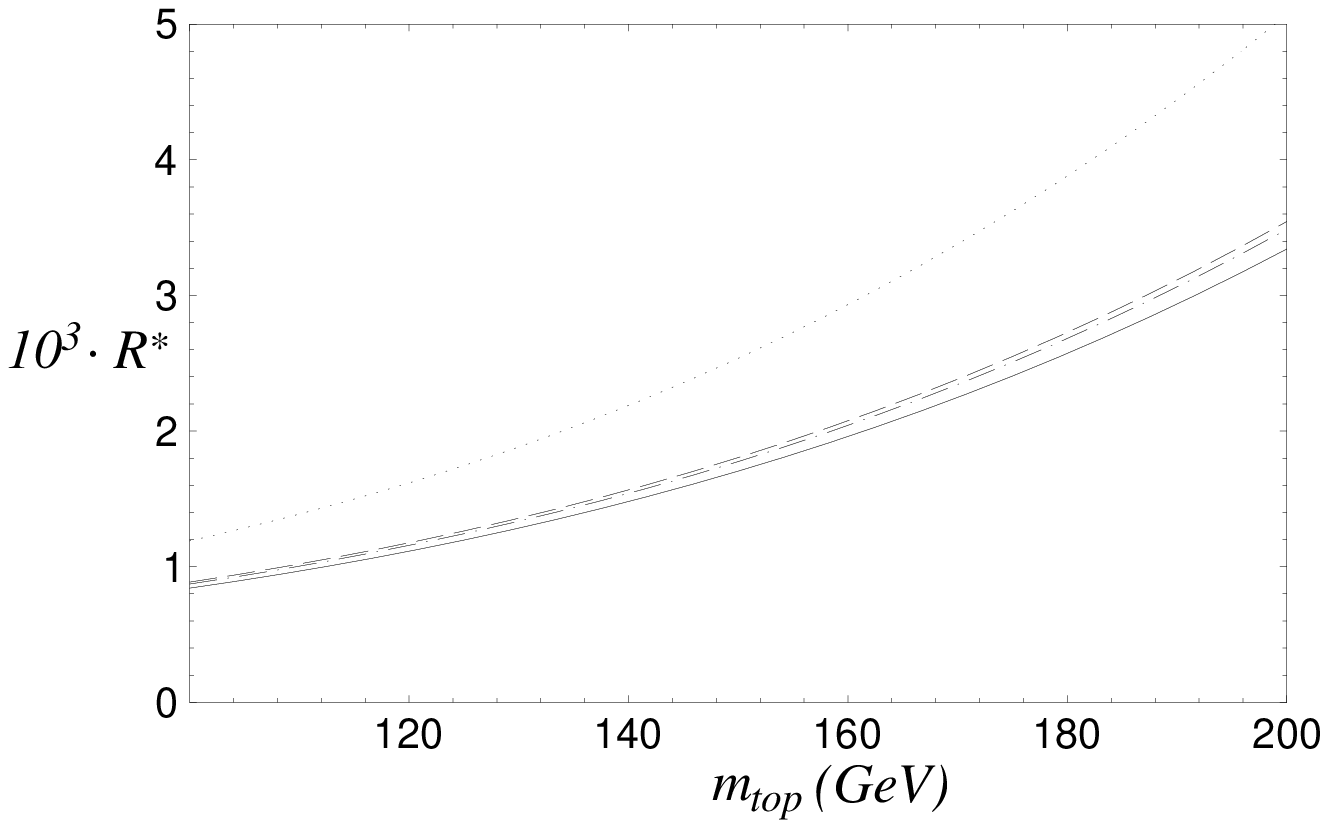}
\end{center}
\caption{Plot of the quantity~$R^{*}$ versus~$m_t$,
for various parametrisations of the
Isgur-Wise function.
}
\label{fig:BKsmt}
\end{figure}

Since we have been calculating ratios of decay rates, it isn't clear
when these processes will become observable at hadron colliders. In
this regard, it is worth pointing out that, in the current Tevatron
run, CDF has roughly a hundred \mathtext{BKspsi}
events, so that the non-resonant \BKsmumu\ process is an order of
magnitude beyond reach.  (From \figref{BKsmt}, $R^*$ is typically of
order $10^{-3}$). In addition, branching ratios involving a
\mathtext{K} in the final state are typically a factor of 3--6 smaller
than the corresponding \Kstar~processes.  However, future planned runs
will begin to put useful bounds on non-Standard Model physics, and
clearly proposed hadron machines such as LHC or SSC will be able to
shed further light on these rare decays.

As can be seen in figures \ref{fig:BKmt}~and~\ref{fig:BKsmt}, substantial
uncertainties are present in our predictions, typically of order 25\% at
fixed top quark mass. These uncertainties stem mainly from not knowing
the momentum dependence of the form factors; in other words, from having to
extrapolate the Isgur-Wise function to larger values
of~\mathtext{v \cdot v^\prime}.
We would like to point out, however, that the procedure of
normalising spectra to the $\psi$~peak (partially motivated by the
constraints of making measurements at hadron colliders) reduces these
extrapolation uncertainties considerably. This can be seen by comparing
\figref{IsgurWise} with the various plots of $R$~and~$R^*$, bearing in
mind that probability distributions are proportional to
\mathtext{\xi^2}.

In fact, our procedure has intentionally been rather crude --- for
example, we have not attempted to include uncertainties in the
Isgur-Wise parameters of equations (\ref{eqn:IWfirst}) --
(\ref{eqn:IWlast}), and consequently we have no quantitative measure
of errors induced by the various parametrisations. Nevertheless,
experimental data can help to reduce and quantify these
errors.  In particular, the long-distance
\mathtext{\psi}~and~\psip\ peaks in the dimuon
spectrum, allow one to measure
\mathtext{B \rightarrow K^{(*)}}~form factors at the corresponding values
of~$q^2$. In our approach, where ratios of
amplitudes are measured, one could use the experimentally measured
value of
\mathtext{
  \Gamma( B \rightarrow K^{(*)} \psi ) \, / \,
  \Gamma( B \rightarrow K^{(*)} \psip )
} to fit the parameters of the Isgur-Wise function experimentally.
This procedure is quite similar to that of
\refcite{Ahmady-Liu}, except that, as mentioned before, we are
avoiding using the overall normalisation of the \mathtext{B
\rightarrow K^{(*)}} form factors, and are reluctant to assume that
\mathtext{\K-\Kstar} spin symmetry holds.

The branching ratios for these processes can be computed from equations
(\ref{eqn:KRate})~and~(\ref{eqn:KStarRate}) via
\begin{equation}
BR\left( B \rightarrow K^{(*)} \psi^{(\prime)} \right)
=
{ \displaystyle 1
  \over 
  BR \left(\psi^{(\prime)} \rightarrow \mup \mum \right) }
\:
BR \left( \left. B \rightarrow K^{(*)} \mup \mum
          \right|_{\psi^{(\prime)}}
   \right),
\end{equation}
where the first factor on the right-hand side is known experimentally.
The second can be calculated numerically, or else, because these two
resonances are so sharply peaked, one can assume the factor~\mathtext{
|A|^2 + |B|^2 } in (\ref{eqn:KRate})~and~(\ref{eqn:KStarRate}) is
dominated by the corresponding term in \eqnref{BreitWigner}.
Integrating (with respect to~$\hat{s}$) over the peak amounts to the
replacement
\begin{equation}
{
  \displaystyle
  1
  \over
  (q^2 - M^2)^2 + M^2 \Gamma^2
}
\rightarrow
{
  \displaystyle
  1
  \over
  M_B^2
  M \Gamma
}
\end{equation}
in which case we have
\begin{equation}
\begin{array}{rcl}
BR\left( B \rightarrow \Kstar \psi^{(\prime)} \right)
&=&
  C^{*}
  M_{\psi^{\prime}} \:
  \Gamma\left( \psi^{\prime} \rightarrow \mup \mum \right) \:
  \xi^2 (y) \:
  \: (1 + y) \: (y^2 - 1)^{1/2} \quad \times
\\[10pt]
& &
  \times
  \left( (\kaps^2 - 8 \kaps + 1)
       + (5 \kaps^2 - 2 \kaps + 5) y
       - 8 \kaps y^2
  \right)
\end{array}
\end{equation}
and
\begin{equation}
BR\left( B \rightarrow K \psi^{(\prime)} \right)
=
  C \:
  M_{\psi^{\prime}} \:
  \Gamma\left( \psi^{\prime} \rightarrow \mup \mum \right) \:
  \xi^2 (y) \:
  (y^2 - 1)^{3/2}
\end{equation}
where $y$~is evaluated at the appropriate kinematic points, and
the constants $C$~and~$C^*$ are the same for both
\mathtext\psi~and~\psip.

Consequently, it is easy to determine, from experimental values,
the ratio of values of the Isgur-Wise functions (for \K~or~\Kstar)
corresponding to the \mathtext{\psi}~and~\psip.
We find
\begin{eqnarray}
\Gamma\left( \BKspsi \right) \, / \,
\Gamma\left( \BKspsip \right)
&=&
3.25 \left(
          \left. \xi_{K^*} \right|_\psi /
          \left. \xi_{K^*} \right|_{\psi^\prime} 
       \right)^2
\\
\Gamma\left( \BKpsi \right) /
\Gamma\left( \BKpsip \right)
&=&
4.52 \left(
          \left. \xi_K \right|_\psi /
          \left. \xi_K \right|_{\psi^\prime} 
       \right)^2.
\end{eqnarray}

Because the decay \mathtext{ \BKpsip } has not
yet been observed, the calculation can only be attempted for \Kstar.
Unfortunately, the errors involved in
the values from \cite{ParticleData},
\begin{eqnarray}
\Gamma\left(B^0 \rightarrow K^{*0} \psi\right)
&=&
(1.3 \pm 0.4) \times 10^{-3}
\\
\Gamma\left(B^0 \rightarrow K^{*0} \psi^\prime \right)
&=&
(1.4 \pm 0.9) \times 10^{-3},
\end{eqnarray}
are quite large, and lead to a ratio of roughly~\mathtext{1 \pm 0.7}.
Models I -- IV give values of 2.28, 2.11, 1.60, 2.18, showing that our
choice of parameters is not too far off.

At this stage, the errors are too large to derive significant
constraints on the parameters of equations
(\ref{eqn:IWfirst})--(\ref{eqn:IWlast}).  Nevertheless, improvements
in the experimental errors, or alternatively a re-analysis of the
existing data with a view to measuring the ratio of rates, could make
this a viable method --- in this case one doesn't have to extrapolate
over too large a range of~$y$.

One might also wonder whether one could use the CLEO measurement
\cite{CLEOpaper} of~\BKsgamma\ to calculate bounds in a similar way.
In this case, there is the problem that this rate could depend on
unknown physics. Even if one assumes the Standard Model holds, one finds
\cite{Ali} that
\begin{equation}
BR (\BKsgamma) \approx (6 \; - \; 10) \times 10^{-4} \: \xi^2
(y_{max} = 3.04),
\end{equation}
where the uncertainty stems from not knowing the top quark mass (we have
assumed that the top quark mass lies between 100 and 200 GeV).
Since in addition the CLEO value (a branching
ratio for \BKsgamma\ of \mathtext{(4.5 \pm 1.5 \pm 0.9) \times 10^{-5}})
contains sizeable errors, we do not believe that reasonable bounds
on parameters will result.

As far as the results for the non-Standard Model
\mathtext{WWZ}~couplings are concerned, it is worth noting that the
contribution of the \mathtext{\Delta g_1^Z}~term is enhanced compared
to that of \mathtext{g_5^Z} by the presence of a logarithmic divergence.
These anomalous vertices appear in graphs very similar to the
ones considered here for radiative corrections to the process
\mathtext{Z \rightarrow b \bar{b}}, where analagous results would
hold (although in that case more than two couplings would contribute).
Consequently, good constraints on these parameters could also
come from $Z$ physics.

\section{Conclusions}
\label{sec:Conclusions}

On the theoretical side, many interesting questions remain open. While
our insistence on ignoring the normalisation of the Isgur-Wise function
might seem overly stringent (especially in the case of the analysis of
\Kstar~modes in Section~\ref{sec:Discussion}) it seems to us to be a
reasonable approach at present. However, we look forward to
improvements in the theoretical picture, which could stem from new
data on decays involving a \mathtext{\psi}~or~\psip\ meson. In
addition, a better understanding of the coefficients of
\mathtext{\bar{s} b \bar{c} c}~operators would be helpful.

Although our analysis has not been very detailed from the experimental
point of view, we believe it constitutes a basis for present and future
searches at hadron colliders. These would of course require a detailed
simulation, taking into account details like backgrounds, and also
fine-tuning our crude cuts in equations
(\ref{eqn:Gamma1})~and~(\ref{eqn:Gamma2}). Nevertheless, we believe that
experiments at future hadron colliders will constitute useful tests of
the Standard Model.

{\Large\bf Acknowledgements}

\nopagebreak
This work was supported by the United States Department of Energy grant
AT03-88ER 40384, Task C. The author would like to thank Thomas M\"uller for
proposing this study, and for useful advice. Discussions with
Roberto Peccei, Carol Anway-Wiese and Fritz de~Jongh have also proved
invaluble.

\vfill
\pagebreak
%
%

\end{document}